\documentclass[sigconf]{acmart}
\AtBeginDocument{%
  \providecommand\BibTeX{{%
    \normalfont B\kern-0.5em{\scshape i\kern-0.25em b}\kern-0.8em\TeX}}}

\copyrightyear{2022} 
\acmYear{2022} 
\setcopyright{rightsretained} 
\acmConference[TechDebt '22]{International Conference on Technical Debt}{May 16--18, 2022}{Pittsburgh, PA, USA}
\acmBooktitle{International Conference on Technical Debt (TechDebt '22), May 16--18, 2022, Pittsburgh, PA, USA}
\acmDOI{10.1145/3524843.3528099}
\acmISBN{978-1-4503-9304-1/22/05}

\usepackage{enumitem}

\begin{document}

\title{MultiDimEr : A Multi-Dimensional bug analyzEr}


\author{Lakmal Silva}
\orcid{0000-0002-5027-9316}
\affiliation{%
  \institution{Blekinge Institute of Technology and Ericsson AB}
  \country{Sweden}
}
\email{lakmal.silva@bth.se}

\author{Michael Unterkalmsteiner}
\orcid{0000-0003-4118-0952}
\affiliation{%
  \institution{Blekinge Institute of Technology}
  \country{Sweden}
}
\email{michael.unterkalmsteiner@bth.se}

\author{Krzysztof Wnuk}
\orcid{0000-0003-3567-9300}
\affiliation{%
  \institution{Blekinge Institute of Technology}
  \country{Sweden}
}
\email{krzysztof.wnuk@bth.se}

\renewcommand{\shortauthors}{Silva et al.}

\begin{abstract}
\textit{\textbf{Background:}} Bugs and bug management consumes a significant amount of time and effort from software development organizations. A reduction in bugs can significantly improve the capacity for new feature development. 
\textit{\textbf{Aims:}} We categorize and visualize dimensions of bug reports to identify accruing technical debt.  This evidence can serve practitioners and decision makers not only as an argumentative basis for steering improvement efforts, but also as a starting point for root cause analysis, reducing overall bug inflow. 
\textit{\textbf{Method:}} We implemented a tool, MultiDimEr, that analyzes and visualizes bug reports. The tool was implemented and evaluated at Ericsson. 
\textit{\textbf{Results:}} We present our preliminary findings using the MultiDimEr for bug analysis, where we successfully identified components generating most of the bugs and  bug trends within certain components.  
\textit{\textbf{Conclusions:}} By analyzing the dimensions provided by MultiDimEr, we show that classifying and visualizing bug reports in different dimensions can stimulate discussions around bug hot spots as well as validating the accuracy of manually entered bug report attributes used in technical debt measurements such as fault slip through. 
\end{abstract}




\maketitle

\section{Introduction}
Software maintenance and technical debt management is a costly process~\cite{dehaghani2013factors}, which is also evident from our experience of building software products at Ericsson. Our agile teams spend between 20\% to 30\% of their time on bug fixing. Hence, bugs impact the feature delivery plans as well as increase the maintenance costs. To further minimize the bug inflow, Ericsson has the ambition to reduce the number of faults slipping through to the customers by 50\% by the year 2022 in a cost-efficient manner.

A natural step to achieve this goal is to analyze historical bug reports. Bug analysis has been a research area for several decades~\cite{rhodes1982automated}. With the advancements in machine learning and natural language processing techniques over the past decade, we see a significant number of studies in areas such as bug report severity prediction~\cite{yang2014towards}, bug report triage~\cite{yang2014towards}, bug report prioritization~\cite{uddin2017survey} and duplicate bug report detection~\cite{jingliang2016data}. Even though these research efforts focus on efficient management of the bugs, there are still improvements required (e.g., classification accuracy, efficiency) on these techniques for industry-wide adaptation~\cite{lee2019systematic}.

We argue that it is beneficial to identify the technical as well as process related reasons that cause bugs in the first place. To achieve our goal of reducing defects, we investigate historical bug reports to identify weak areas of a software system to take preventive measures that reduce faults slipping to the customer. Staron et al. suggested a method to quantify and visualize code stability using heat maps~\cite{staron2013measuring}. Their study concluded that the visualization method could effectively identify error-prone components and trigger software quality improvements. Further, their study hinted at adding multiple dimensions to the metrics as future work. We believe that such multi-dimensional views can stimulate improvement discussions based on the identified trends in reported bugs. Our study proposes a novel approach by introducing multidimensional analysis into the bug analysis.
The main contributions of this paper are:
\begin{enumerate}[label=(\alph*)]
\item A proposal of ten dimensions that characterise bug reports which can be used to analyze trends.
\item Lessons learned from building a bug analysis tool in an industrial setting.
\item Early feedback from the users of the tool.
\end{enumerate}

The remainder of the paper is structured as follows. Section~\ref{sec:rw} provides an overview of prior research on bug analysis and visualization tools, while Section~\ref{sec:redesign} describes the research design. Section~\ref{sec:mde} provides the concepts and motivations behind the tool MultiDimEr and its implementation details. We present our preliminary findings from an empirical analysis in Section~\ref{sec:pr} and conclude with an outline for our future work in Section~\ref{sec:cf}.

\section{Related Work}\label{sec:rw}
D'Ambros et al.~\cite{d2007bug} proposed two visualizations of bug distributions (over system components and time) and bug life cycles. The  combination of graphical views as a means to detect and analyze hidden patterns within bug reports was also suggested by Knab et al.~\cite{knab2009interactive}. Their tool focuses on bug report attributes such as estimated effort, actual effort, priority, and ownership. 

With the tool BugMaps~\cite{hora2012bug}, Hora et al. extracted information about bugs from bug tracking systems for visualization. The tool provided a historical view and a snapshot view of the bugs by linking them with source code classes.

Chang et al. proposed a defect prediction method~\cite{chang2009integrating} using association mining techniques to automate root cause analysis to prevent bugs.

To identify version-related bug patterns, Sun et al. introduced a technique by finding similar bugs from bug repositories and combining the bug pattern with a ranked code snippet~\cite{sun2019bug}.

Cervantes et al. introduced Archinaut~\cite{cervantes2020software}, a tool that is geared towards understanding software architecture and supporting architectural technical debt (ATD) management. The tool also provides a visualization of the bug density on individual source code files.

MultiDimEr is different from these approaches as we employ a multi-dimensional approach on the bug report data to obtain a differentiated view of the product's status and evolution. Our aim is to identify bug reporting trends to steer improvements and take preventive measures to minimize the accumulation of technical debt. Classification and visualization of bug distribution over architectural components is one of the unique features of MultiDimEr.

\section{Research Design}\label{sec:redesign}

The goal of our research is to evaluate the idea of analyzing historical bug reports by visualizing their occurrence along various dimensions (discussed in Section~\ref{sec:mde}). We conjecture that this allows us to identify trends, which can deliver insights on the root causes for defect introduction. MultiDimEr is a prototype we built as a base for this analytical work. We developed MultiDimEr in multiple iterations, based on feedback from a bug management forum at Ericsson. We conducted analysis sessions together with practitioners, aiming at answering the following research questions:

\begin{enumerate}[label=RQ\arabic*]
  \item \label{itm:components} What components within the system are causing the most bugs?
  \item \label{itm:patterns} What trends can be observed from components generating most of the bugs?
  \item \label{itm:validation} To what extent can MultiDimEr be used to validate the accuracy of fault slip through analysis?
  \item \label{itm:lessons} What are the lessons learned when implementing MultiDimEr in an industrial context?
  \item \label{itm:benefits} How useful is MultiDimEr from a technical debt management point of view?
\end{enumerate}  

\ref{itm:components} can be answered by studying the visualization provided by MultiDimEr. To answer \ref{itm:patterns}, MultiDimEr provides the list of bug reports for the selected component along with crucial bug report attributes. We conducted a pilot study to analyze and understand how feasible it is to answer  \ref{itm:components} and \ref{itm:patterns} with MultiDimEr. The results are reported in Section~\ref{sec:pr}. It should be noted that the analysis process is currently ongoing at Ericsson; hence only the available results at the time of writing this paper are presented. 

\section{M\lowercase{ulti}D\lowercase{im}E\lowercase{r}}\label{sec:mde}
\graphicspath{ {./images/} }

This section provides the concepts behind the MultiDimEr and its implementation. We emphasize on the visualization aspect since it is a powerful way to convey the results. The main idea behind MultiDimEr is to classify bugs based on different dimensions. We considered how bug reports are distributed over: (1) architectural components, (2) source code files, (3) action that has been taken concerning a bug (answer code), (4) the bug reporting countries, (5) customers, (6) bug detection phase, (7) customer-facing documents, (8) software revisions, (9) bug severity, (10) bug report status.

The bug analysis is scheduled to run every 12 hours, providing up to date analysis. We also implemented an on demand analysis based on a selected release/releases within a certain time frame. This is beneficial for the management since they are required to report bug analysis on a monthly basis. Currently, this analysis is performed manually, only focusing on customer bug reports. The shorter feedback loops from MultiDimEr facilitates routine bug analysis in the ten dimensions listed above, that significantly improves the bug analysis and reporting process. 
The following sub-sections describe the chosen dimensions and the motivations behind the choices. 

\subsection{Bug distribution over architectural components} \label{multirelease_distrib}
\begin{figure}[h]
  \includegraphics[width=8cm ,height=2cm]{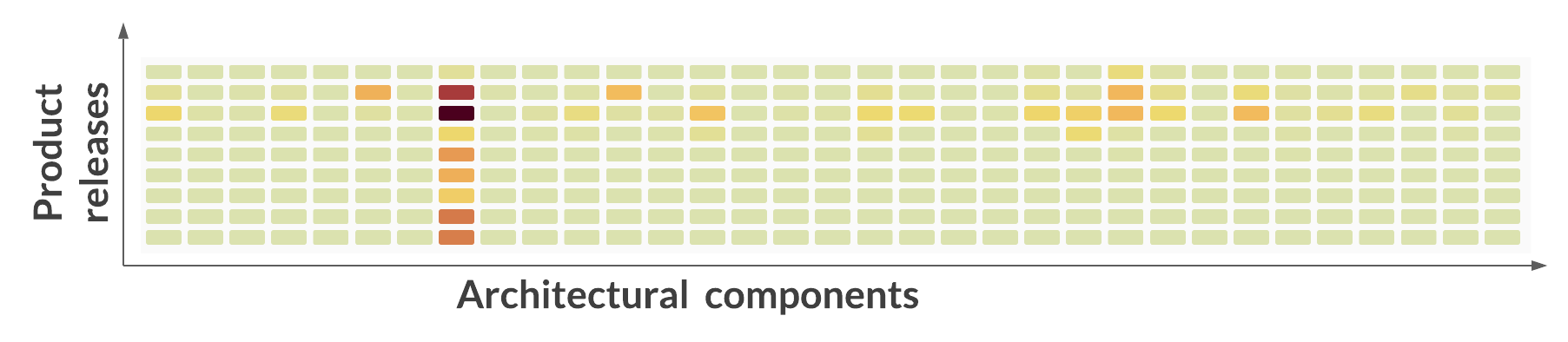}
  \caption{Heat map of bug distribution over Architectural components over releases }\label{heatmap}
\end{figure}

MultiDimEr visualize the bug distribution per architectural component over multiple releases, as shown in Figure~\ref{heatmap}. The architectural components and their granularity were agreed upon with the Architecture team of the product under investigation. They represent logical functional blocks of the software architecture and are associated with the source code repositories and folder structures that are being used for bug report mapping as described in~\ref{impl}. The x-axis of the heatmap represents different Architectural components, while the y-axis represents the product releases from oldest to the newest release. The darker the element, the higher the bug density of a component. By visual inspection, we can see that certain components have an increasing bug density as the system evolves. We plan to investigate this observation in a future study to understand the reasons for the increased number of bugs in specific components.

An analysis of the bug distribution within a single release helps to narrow down components to be prioritized for further investigations.  

NOTE: The Product release names and the Architectural component names have been removed due to confidentiality agreements. 

\subsection{Bug distribution over source code} \label{srccode_distrib}
\begin{figure}[h]
  \includegraphics[width=9cm ,height=1.25cm]{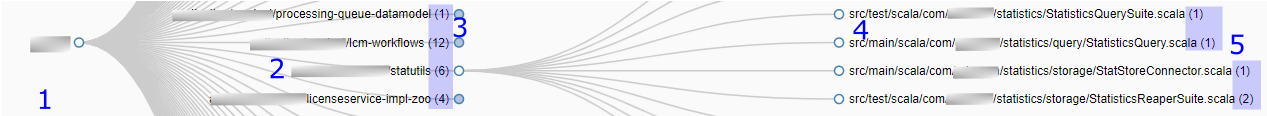}
  \caption{Bug distribution over source code files }\label{sourcode}
\end{figure}

Mapping bugs to source code can reveal hot spots within the source code. The idea is similar to Archinaut's~\cite{cervantes2020software} bug visualization. The source code is visualized as an interactive collapsible tree with accumulated bug counts on each branch as shown in Figure~\ref{sourcode}. The graph starts with the product name~(1). Clicking on the product level expands the source code repositories~(2) with the corresponding bug counts~(3) displayed on branches. Based on the repository bug densities, users can traverse to the individual source files~(4), which shows the bug density on files~(5). This allows to narrow down error-prone files. In the future we plan to triangulate the error-prone files detected by MultiDimEr with the reports from static code analysis tools to identify relationships between bug densities and coding violations.

\subsection{Bug distribution over answer codes}
The bug management system contains an attribute called ``answer code'', filled by designers when answering a bug report. The answer code indicates which actions will be taken on the bug. An answer code belongs to one of one three categories, ``Already Corrected'', ``Will be corrected'', or ``No Action''. The answer code is a vital attribute used in process measurements such as Fault Slip Through (FST)~\cite{damm2006faults}. This visualization helps to validating the correctness of the manually entered attribute as we have shown in \ref{sec:pr}.

\subsection{Bug distribution over bug reporting countries}
Ericsson is selling and supporting products in various countries that require customization to local rules and regulations. We added the distribution of bugs over bug originating countries to determine whether there are bug reporting trends based on different regions. A deeper analysis in this dimension could provide insights into whether certain countries use a particular feature set or adaptations that contribute to more bugs than other countries.  

\subsection{Bug distribution over customers}
Customer satisfaction is at the core of a successful business. The number of bugs reported by customers can be considered an indicator of how the customers perceive the system. A visualization of the bugs on specific customers provides a quick overview of customer-specific issues so that the development units can further analyze the error-prone components within a system.

\subsection{Bug distribution over bug detection phase}
The bug detection phase such as function testing, system testing, integration testing and customer testing, is an important parameter that is being used in FST analysis. The visualization of bug reports distributed over different test phases allows us to validate if this parameter has been set correctly by the bug reporters. 

\subsection{Bug distribution over customer-facing documents} 
Customer-facing documents are part of the product and generate bug reports when incorrect information has been detected. As there are various documents, such as installation guides, configuration guides, and interface documents, it is necessary to visualize the bug distribution over different documents to prioritize document improvements.

\subsection{Bug distribution over severity}
The classification of bugs based on severity alone is not that interesting. However, combining it with other dimensions can provide interesting and actionable insights. We plan to investigate different bug dimension combinations in a future study, for instance:

\begin{enumerate}[label=(\alph*)]
\item bug severity distribution per architecture components
\item bug severity distribution on bug detection phase
\end{enumerate}

\subsection{Visualization}
\begin{figure}[h]
  \includegraphics[width=8cm ,height=2cm]{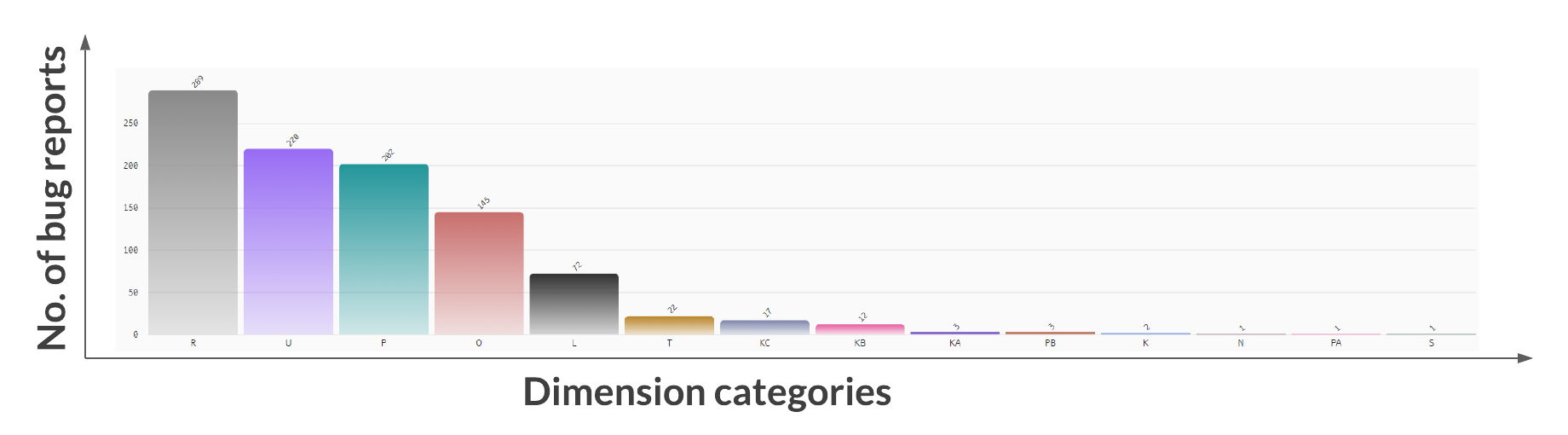}
  \caption{Visualization of bug distribution over a selected dimensions}\label{histograms}
\end{figure}

Except for the visualization of the dimensions described in Sections~\ref{multirelease_distrib} and \ref{srccode_distrib}, the bug distributions are visualized as bar charts to contrast frequencies of components of a selected dimension, as illustrated in  Figure~\ref{histograms}. Different components within a chosen dimension are distributed over the x-axis in descending order, starting with the component with the highest number of bugs, while the y-axis shows the number of bugs. 
The plots are interactive, so it is possible to see the bug list contributing to the bug count on different plots by clicking on the bug counts. This approach facilitates getting a differentiated view of the types of bugs within a specific component. 

MultiDimEr also provides hyperlinks to access the raw bug report from the bug management system. This was added to improve the user experience and find additional details that might have not yet been incorporated into MultiDimEr.

\subsection{Implementation} \label{impl}
This section provides an overview and the development journey of the MultiDimEr tool. The system constitutes of a back-end and a front-end, packaged as Docker\footnote{https://www.docker.com/} containers and deployed into a Kubernetes\footnote{https://kubernetes.io/} container orchestration system. The back-end was implemented using  Django\footnote{https://www.djangoproject.com/}, a Python web framework, and the Python data analysis and manipulation tool Pandas\footnote{https://pandas.pydata.org/}, while the front end was implemented using the Angular web framework\footnote{https://angular.io/}. The back-end exposes a set of REpresentational State Transfer (REST) Application Programming Interfaces (APIs) for triggering configuration updates, starting the bug analysis process, and retrieving processed data. The front-end consumes these interfaces to implement an interactive Graphical user interface (GUI).  

\begin{figure}[h]
  \includegraphics[width=9cm ,height=5cm]{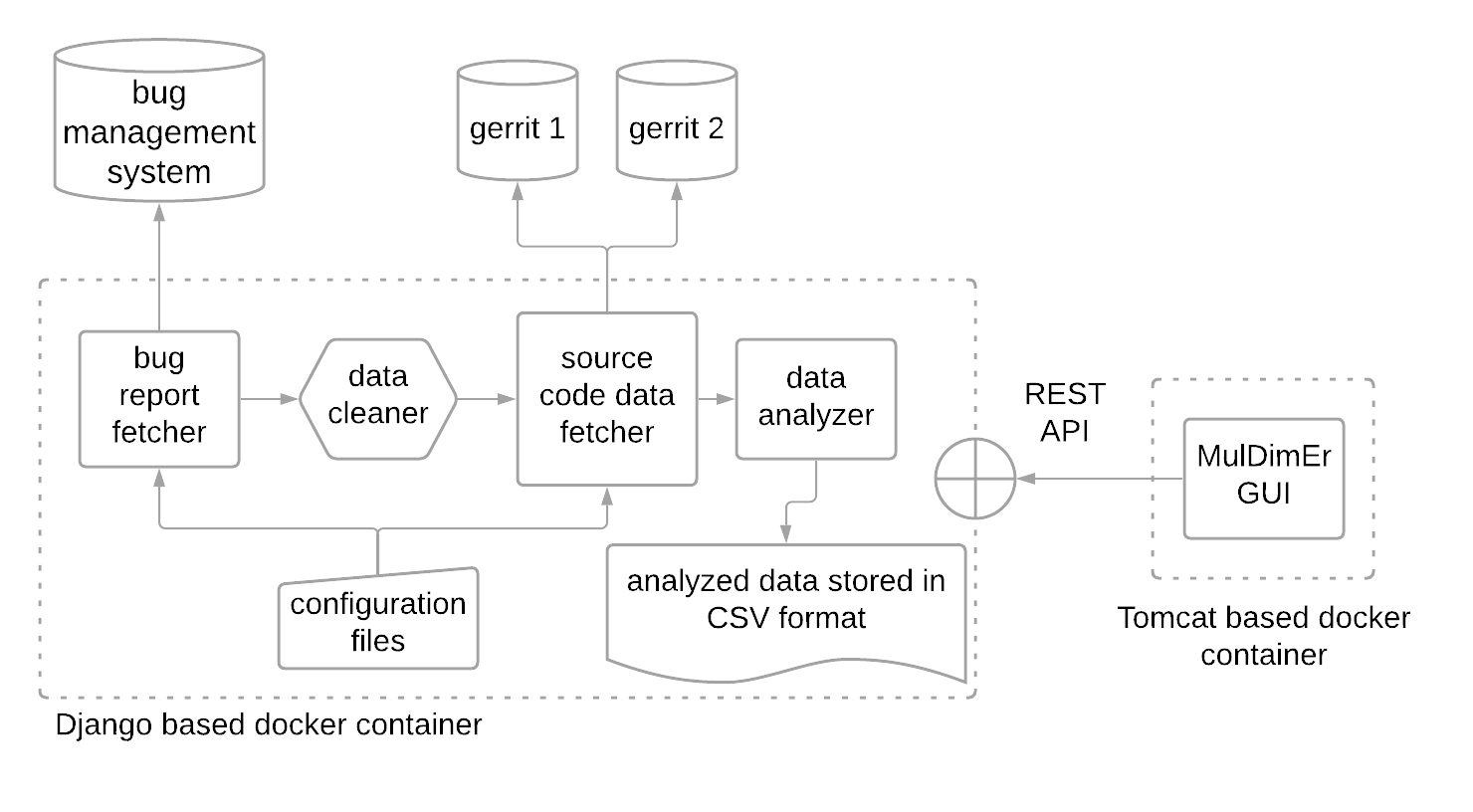}
  \caption{MultiDimEr Architecture}\label{architecture}
\end{figure}

Figure~\ref{architecture} depicts the high-level architecture of MultiDimEr. The bug report analysis process is started from the GUI or the REST API by providing the product number(s) and the time frame for bug report collection. The ``bug report fetcher'' module uses this information and configured credentials to retrieve the selected product's bug reports within the chosen time frame. Once the data is retrieved, the ``data cleaner'' module processes the answer section of the bug report to extract commitids/changeids. Identifier extraction was a challenging task as the developers used different ways to represent the Ids. We used a regular expression matching mechanism with different patterns to extract the identifiers. 

Once the identifiers have been extracted, they are sent over to the ``source code data fetcher'' module. This module uses credentials and the Gerrit identifiers from the configuration file to query for commit details (the repository name, changed file set) over the Gerrit REST API. In the ``data analyzer'' module, data collected from the previous modules are processed, consolidated, and written to a comma-separated values (CSV) file, that can be further processed by tools other than the MultiDimEr if necessary. 

One of the ``data analyzer'' modules' essential tasks is to map bug reports to the architecture components. Bug report mapping to architecture components is performed based on two dictionaries and the Gerrit system's information. One dictionary maps the git repository name to an architecture component. The second dictionary maps file paths of a given commitid/changid to an architecture component. It was necessary to create these two types of dictionaries since the source code management repositories have evolved from a monolithic source repository to modular repositories during the effort of decomposing the monolithic software system into modules/microservices. 

\section{Results}\label{sec:pr}
In this section, we report our findings and answer the research questions.

\paragraph{\ref{itm:components}: What components within the system are generating the most number of bugs?}
Most bugs resulted in updates to customer-facing documents, followed by platform (functionality associated with installation and upgrades) related bugs. 

\paragraph{\ref{itm:patterns}: What trends can be observed from components generating most of the bugs?}
The visualization of bug distribution over customer-facing documents facilitated identifying trends in document defects that can be used to in improvement strategies:

\begin{enumerate}
  \item Lack of attention on the impact of customer-facing documents when updating functionality.
  \item Lack of focus on the knowledge of the end-users. Documents were written by the developers without considering the end-users' technical background.
  \item Incorrect commands were present in the documents due to lack of document verification.
\end{enumerate}

An investigation into an asynchronous interface component revealed bug trends due to poor handling of message queues, notifications and robustness. A study has been started to rework this component to improve its robustness.

Bug distribution over customer view revealed that around 83\% of the bugs were detected internally, which indicates good testing processes already in place. We are currently investigating the bug trends within the 17\% of actual customer bugs.

\paragraph{\ref{itm:validation}: To what extent can MultiDimEr be used to validate the accuracy of fault slip through analysis?}
MultiDimEr detected that a significant number of bugs were answered with codes belonging to the answer code group ``Already Corrected''. From our experience, it was very unlikely that such a large number of bugs have been detected and fixed before a bug report has been created. The root cause was the developers were using incorrect answer codes when answering the bug reports. Further analysis also showed that most of the developers that used the incorrect codes were newcomers to the organization. This was a significant finding as the FST analysis was providing inaccurate results due to the incorrect code usage. A sensible improvement is to update the training material related to reporting bug reports and possibly improve the bug report tooling such that manual mis-classification errors occur less frequently.

\paragraph{\ref{itm:lessons}: What are the lessons learned when implementing MultiDimEr in an industrial context?}  We observed that the source code had been moved around into different repositories over the years, without any documented references to such restructurings. However, MultiDimEr detected this anomaly as it did not manage to get data for some of the extracted commit identifiers. Therefore, we recommend to document the major source code structural changes such as decomposing repositories, moving and renaming of repositories, to trace such significant changes. This simplifies tool development related to data extraction from source code repositories over a longer period of time.

Compared to mapping bugs directly to source code files, as suggested by Archinaut~\cite{cervantes2020software}, our approach was to map bug reports to architecturally significant functional components, a higher abstraction level than the source code. This helped us to group sufficient amounts of bugs to identify bug trends. It is recommended to use release agnostic names when analyzing bug distributions over multiple releases as the actual components names may change during system evolution.

The implementation became more complex as we had to process free text to find git changeids and commitids from the bug reports. It is worthwhile enforcing a separate attribute for the commitid/changeid when answering bug reports. That could significantly simplify tool development. 
\paragraph{\ref{itm:benefits}: How useful is MultiDimEr from a technical debt management point of view?}

To answer \ref{itm:benefits}, we interviewed two practitioners that are actively using MultiDimEr for bug analysis. Summarizing their feedback:

\begin{itemize}
\item The visualization is simple, which makes users to clearly see the bug distributions in different dimensions.
\item The bug distribution over architecture components clearly visualize the problematic areas within the system, which were unknown without MultiDimEr, and provides a starting point to understand weak areas within the system.
\item Bug report mapping from source code to architecture components is appreciated, since we do not need to change the current ways of working in the bug management forum or in development teams.
\item The system improvements suggested by testing teams never get prioritized by the product owners over new feature implementation. MultiDimEr analysis provides evidence for motivating why improvements should be prioritized in error-prone areas.
\item Bug distribution over customers helps us to understand the features they use, what features are of low quality, if the bug reports originating from the base features or customer specific adaptations, and if all customers are experiencing issues with similar areas. 
\item Different dimensions provided by MultiDimEr make it usable by multiple groups such as testers, developers, architects, support personnel, and customer units. For instance, bug distribution is interesting for architects to understand improvement areas and the component responsible teams can use it for refactoring initiatives and improvements.
\item MultiDimEr provides aggregated information compared to the current approach of using different tools to collect data and manual analysis. Due to the amount of time required for manual analysis, it is often omitted. 
\item FST analysis is conducted to improve test processes. However, the FST is seldom performed as the analysis needs to be carried out manually, consuming a lot of time. MultiDimEr provides us the opportunity to carry out FST analysis more efficiently and more frequently.
\item The CSV file based result output provides more opportunities for us to combine additional dimensions/filters that are not yet implemented in the GUI.
\item The number of sentences present in the bug report observations and answer sections gives an indication of how well the bugs have been described and how well the developers documented fixing it. 
\end{itemize}

\section{Conclusion and future work }\label{sec:cf}
This study presented MultiDimEr, an analysis and visualization tool to classify bug reports in different dimensions. We piloted the tool in an industrial context at Ericsson, trying to identify bug trends and allow root cause analysis. Since this is an ongoing process, so far we analyzed three dimensions out of the ten dimensions provided by MultiDimEr, the bug distribution over architectural components within a single release, customer facing documents and the answer codes. From the architectural component dimension, we discovered that customer-facing documents generated most of the bug reports within the investigated product release. 

The answer code dimension revealed that a significant number of bug reports were not answered with correct codes. Since the answer codes are used for evaluating the process efficiencies, incorrect values mislead the evaluations. 
We plan to study the implications of the implemented actions based on our analysis to reduce document-related bugs in the coming months. Additionally, we plan to examine the reasons for bug distribution variations over architectural components as the system evolves, how to combine dimensions provided by MultiDimEr, and how different stakeholders such as software architects, developers, and testers can utilize insights from MultiDimEr for improvements.

\bibliographystyle{ACM-Reference-Format}
\bibliography{ref}

\end{document}